# Monte Carlo simulation of carbon ion radiotherapy for Human Eye

Cheng-guo Pang[1,2] (庞成果),You-wu Su[2] (苏有武),Ze-en Yao[1](姚泽恩),Jun-kui Xu[1,2](徐俊奎)，Wu-yuan Li（李武元），Jiao Yuan[2]（袁娇）

1,School of Nuclear Science and Technology, Lanzhou University, Lanzhou 730000, China

2,Institute of Modern Physics, Chinese Academy of Sciences, Lanzhou 730000, China

**Abstract:** Carbon ion is the mostly common used particle in heavy ion radiotherapy. In this paper, carbon ion dose in tumor treatment for human eye was simulated with FLUKA code, 80 MeV/u carbon beam was irradiated into the human eye from two directions, The first is from the lateral-forward direction which was a typical therapeutic condition, maximum dose was deposited in the tumor volume. The second one was that beam irradiated into eyes from the forward direction which may cause certain medical accident. The calculated results are compared with other reports. The agreement indicates that this method can be used for treatment plan in heavy ion radiotherapy.

## 1.Introduction

Uveal melanoma is the common tumor in the eyes of adults and children, which accounting for about 12% of all melanomas[1], Its morbidity is just under the retinoblastoma. In terms of this kind of tumors, the usual therapy is ophthalmectomy, transpupillary thermotherapy, brachytherapy and etc. However, those methods have the same risk that the tumor may be spread to other place or cause other serious consequence.

For uveal melanoma treatment, heavy ions therapy, especially using carbon ions have many advantages: firstly, carbon ions deposit their maximum energy density at the end of their track which is the so called Bragg peak. The comparison for the Bragg peak of different particles is shown in Figure 1[2]; secondly, carbon ions can easily be formed as narrow focused and scanning pencil beams of variable penetration depth; which are very important because the critical organs necessary for eyesight are located very close; thirdly, the RBE value of carbon ion is higher than mostly common radiation; lastly, The location where the dose is deposited by carbon ions can be determined by means of the online positron emission tomography.

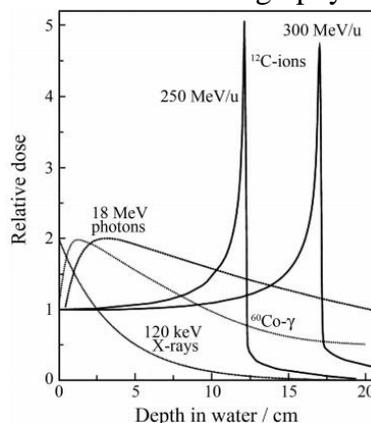

Figure 1:The comparison for dose- depth distribution in water of X-rays, $^{60}$Co- γ rays, high energy photons and 250 MeV/u、300 MeV/u carbon ions

FLUKA is a general purpose tool for calculations of particle transport and interactions with matter, which is jointly developed by the European Laboratory for Particle Physics (CERN)[3], and the Italian National Institute for Nuclear Physics (INFN). Sixty different particles plus heavy ions can be transported by this code over a wide energy range, so it is suitable for calculation heavy ion dose in radiotherapy. The aim of this paper is to develop a simple model of human eye to estimate dose delivered from carbon ions radiotherapy. The calculated dose included that due to carbon ions and secondary particles.

## 2.Simulation of energy deposition in water

To simulate the dose distribution of human eye, First of all, the range of 80MeV/u,85MeV/u,90MeV/u,95MeV/u and 100MeV/u carbon ions in water was calculated with SRIM2013. the result is shown in Table 1.

Table 1: the range of different energy carbon irradiated into water

| Energy(MeV/u) | Range(cm) |
| --- | --- |
| 80 | 1.73 |
| 85 | 1.94 |
| 90 | 2.14 |
| 95 | 2.36 |
| 100 | 2.59 |

Due to the small size of human eye which radius is about 1.3 cm, according to the range of carbon ions in tissue, the energy of 80MeV/u for carbon ions is enough for eye radiotherapy.

To verify whether the FLUKA code is suitable for heavy ions dose simulation in radiotherapy, the dose deposition of carbon ions with energy of 80MeV/u in water was calculated beforehand. The calculated results are compared with the relative ionization energy-depth curve(fig.2)[4], good agreement indicate that FLUKA is competent for simulating heavy ions dose in radiotherapy.

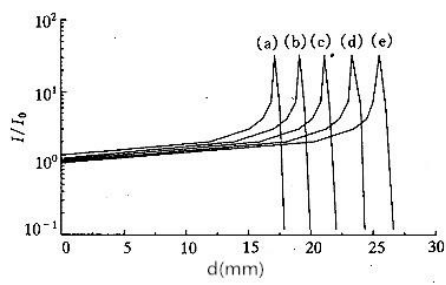
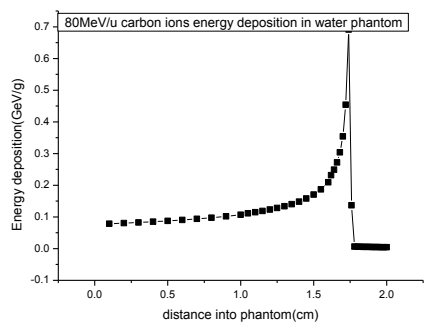

Figure2:Relative ionization energy-depth curve(left,a-80MeV/u,b-85MeV/u,c-90MeV /u, d-95MeV/u,e-100MeV/u), FLUKA simulation results (right)

## 3.FLUKA simulation

The two dimensional rendering of the eye used for the FLUKA modeling is shown in Figure 3. On account of many uveal melanomas appear in the choroid and sclera structure of the eye, many critical structures of eye are radiosensitive such as the lens, the cornea etc. which are located very close, cataract is one of the deterministic effect of radiation with relatively low threshold dose, and radiation exposure to the cornea can cause the structure to become opaque that lead to blind, this should be considered carefully in treatment, and those critical area should be distinguished in the model

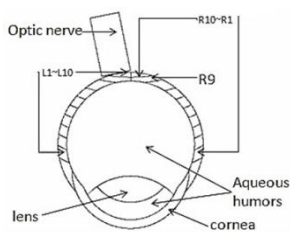

Figure 3: FLUKA simulation model of human left eye

Overall, the model was constructed with a set of concentric spheres. The optic nerve is simulated as a cylinder appropriately offset at the posterior of the eye. Assuming cancerous tumor located in the volume R9 in Figure 3, The L1 to L10 and the R1 to R10 is ten dispersed volumes of left and right side of the eye model respectively. two cases which the direction of carbon beam is from lateral forward and forward direction respectively is simulated as shown in Figure.4.

Material compositions of the eye were adapted from the ICRU Report 46[6]. This report addressed various tissues groups in the body and defined their elemental composition and density for purposes of radiation dosmetry. The lens of the eye is addressed directly in the MCNPX$^{TM}$ user's manual version 2.5.0[7], Recent studies have indicated the vitreous and the anterior humors have characteristics similar to the properties of lymph outlined in ICRU 46, therefore, the composition of the vitreous and anterior is assumed to be the same as the lymph, the choroid and sclera are considered to be soft tissue, and the composition of optic nerve was assumed to be the same as the rat's [8] .

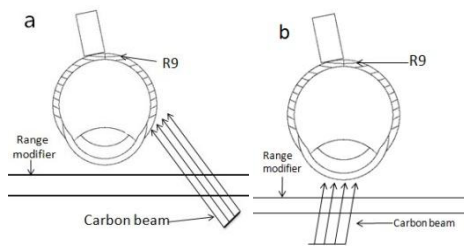

Figure 4: Illustrations of lateral forward direction treatment(a) and forward direction treatment(b) geometry

## 4. Results and analyses

The relative error(R) of the simulation results is mainly caused by the statistical fluctuation which can be control via changing the historical number of particles in simulation. The relative error for two cases treatment program were 0.0056 and 0.0038 respectively.

### 4.1 The lateral forward direction treatment program

According to some recent studies , the lateral-forward direction therapeutic dose for uveal melanoma in the cancerous tumor volume R9 is about 50Gy spread over four fractions, which translates into four treatments of 12.5Gy delivered to the patient. The simulation results for the lateral-forward direction treatment program are found in Table 2. According to the results, the cancerous tumor volume R9 get the max dose (12.5019Gy per fraction, 50.0078Gy in total ) while minimizing dose else where. The dose of left side at least three orders of magnitude lower than right side. Total dose to the optic nerve was only about 6.8764 Gy, This is within acceptable the limit of 10Gy that optic nerve for radiotherapy[9], For each fraction the cornea received less than 0.1 Gy, well within the acceptable limit of 15 Gy[10] , For the lens of the eye, special effort is made in radiotherapy to keep doses at an acceptable limit: usually less than 8Gy , In this simulation, cumulative dose to the lens of the eye from four fractions was only 0.0024Gy.

Table 2: Dose distribution for lateral-forward direction treatment program

| Dose volume | Dose per Fraction(Gy) | Total Dose(Gy) |
| --- | --- | --- |
| Cornea | 0.0134 | 0.0534 |
| Anterior humor | 0.0005 | 0.0019 |
| Lens | 0.0006 | 0.0024 |
| Vitreous humor | 1.8266 | 7.3064 |
| Optic NERVE | 1.7191 | 6.8764 |
| R1 | 2.1192 | 8.4768 |
| R2 | 2.2419 | 8.9679 |
| R3 | 2.2137 | 8.8549 |
| R4 | 2.1032 | 8.4128 |
| R5 | 1.9426 | 7.7705 |
| R6 | 2.0001 | 8.0005 |
| R7 | 3.3672 | 13.4689 |
| R8 | 10.5536 | 42.2142 |
| R9 | 12.5019 | 50.0078 |
| R10 | 7.5663 | 30.2654 |
| L1 | 0.0003 | 0.0013 |
| L2 | 0.0004 | 0.0017 |
| L3 | 0.0007 | 0.0026 |
| L4 | 0.0011 | 0.0039 |
| L5 | 0.0017 | 0.0068 |
| L6 | 0.0034 | 0.0136 |
| L7 | 0.0108 | 0.0431 |
| L8 | 0.0839 | 0.3358 |
| L9 | 0.2041 | 0.8164 |
| L10 | 0.2357 | 0.9427 |

In this paper, we used a proper card USRBIN of FLUKA to score the dose distribution of each section for the eye model. Figure 5 and Figure 6 revealed the results of several sections in which the color more deep, the dose more high.

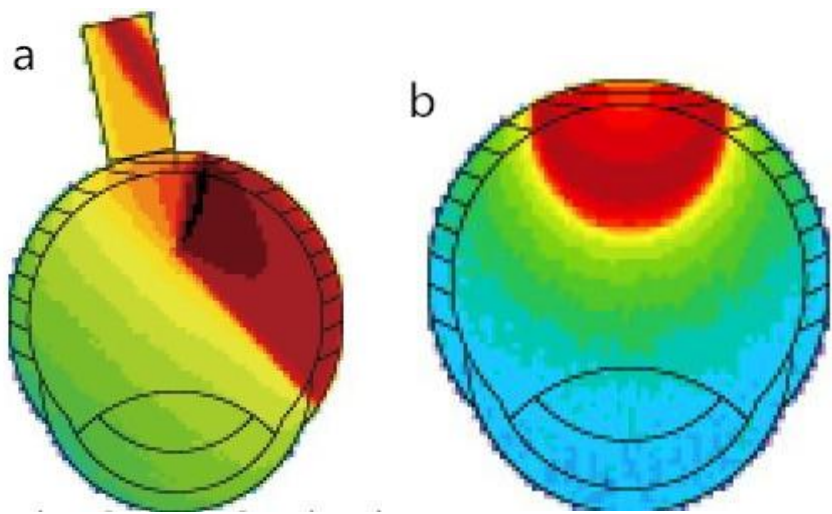

Figure 5:Dose distribution for section when y=0cm(a), and x=0cm(b)

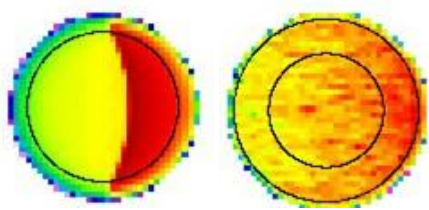

Figure6:Dose distribution for section at z=4.6cm(left), and z=2.3cm(right)

## 4.2 The forward direction treatment program

For the forward direction scenario, we mimicked a patient gazing into the beam during treatment. The simulation results are shown in Table 3.

In accordance with the Table 3, the majority of the energy is deposited in the lens, anterior humor, vitreous humor and cornea. However, the dose of the cancerous volume is almost zero. In this manner, just about 5.2411Gy would be delivered to the cornea which is well within the limits of 15Gy and, if this configuration occurred for the duration of treatment, the patient would suffer over 25Gy to the lens, compared to the accepted tolerable dose of 8Gy, one would expect severe visual loss due to the lens becoming opaque in this treatment. The dose distribution for section of the eye model is illustrated in Figure 7 and Figure 8.

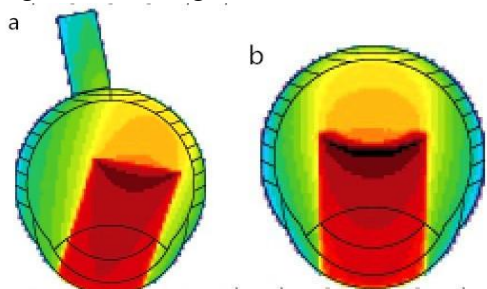

Figure7: Dose distribution for section when y=0cm(a), and x=0cm(b)

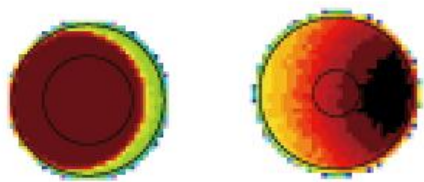

Figure 8: Dose distribution for section at z=4.6cm(left), and z=2.3cm(right)

Table 3: Dose distribution for forward direction treatment program

| Dose volume | Dose per Fraction(Gy) | Total Dose(Gy) |
|---|---|---|
| Cornea | 1.3103 | 5.2411 |
| Anterior humor | 4.1885 | 16.7538 |
| Lens | 6.2868 | 25.1743 |
| Vitreous humor | 1.9569 | 7.8276 |
| Optic NERVE | 0.0028 | 0.0113 |
| R1 | 0.0028 | 0.0113 |
| R2 | 0.0039 | 0.0158 |
| R3 | 0.0068 | 0.0273 |
| R4 | 0.0156 | 0.0624 |
| R5 | 0.0352 | 0.1409 |
| R6 | 0.0551 | 0.2205 |
| R7 | 0.0678 | 0.2713 |
| R8 | 0.0747 | 0.2988 |
| R9 | 0.0695 | 0.2781 |
| R10 | 0.0568 | 0.2273 |
| L1 | 0.0007 | 0.0029 |
| L2 | 0.0006 | 0.0025 |
| L3 | 0.0006 | 0.0025 |
| L4 | 0.0007 | 0.0029 |
| L5 | 0.0009 | 0.0037 |
| L6 | 0.0013 | 0.0053 |
| L7 | 0.0021 | 0.0085 |
| L8 | 0.0049 | 0.0019 |
| L9 | 0.0127 | 0.0508 |
| L10 | 0.0179 | 0.0719 |

## 5 Conclusion

The objective of this paper was to develop a model of human eye using the computer code FLUKA that estimates dose delivered during radiotherapy. On the basis of the simulation results. we can draw the following conclusions:

1) The Monte-Carlo simulation of heavy ions radiotherapy is a capable method using to formulate the patient planning

2) Results for the lateral forward direction treatment program indicate it could be regarded as a typical treatment program for the dose to the tumor volume was at therapeutic levels and at the same time, doses to the cornea, lens and optic nerve were within acceptable limits.

3) For the forward direction treatment program, the result was a large dose to the lens, but the tumor did not receive any dose in this configuration. Compared with the result of the lateral forward direction treatment program, this scenario should be avoided in reality radiotherapy. In the long term, the result of this case could be used to dose distribution reconstruction in a medical negligence.

There are also some problems should be solve in the future. Greater details could be incorporated into the current of the eye-effectively expanding the types of cancerous tumors which might be modeled. Regions outside the eye were neglected in the creation of this model. Inside the eye, greater accuracy could also be attained by adding more detail. Within the current eye model it was not be possible to simulate these types of tumors and their treatments. Likewise, greater resolution could be obtained by differentiating between organs located closely together in the eye .